\documentclass[a4paper]{jpconf}
\usepackage[utf8]{inputenc}
\usepackage[english]{babel}
\usepackage{amsmath}
\usepackage{amssymb}
\usepackage{graphicx}
\usepackage{listings}
\usepackage{color}
\usepackage[binary-units=true]{siunitx}
\usepackage{units}
\usepackage{subcaption}
\usepackage{url}
\usepackage{hyperref}
\usepackage{cite}
\usepackage[colorinlistoftodos,prependcaption,textsize=tiny]{todonotes}
\usepackage{xargs}
\usepackage{float}
\usepackage{listings}
\usepackage{tikz}

\begin{document}
\title{Photoproduction of the $\Lambda$(1520) hyperon with a \SI{9}{\GeV} photon beam at GlueX}

\author{Peter Pauli for the GlueX Collaboration}

\address{SUPA School of Physics and Astronomy, University of Glasgow, UK}

\ead{p.pauli.1@research.gla.ac.uk}

\begin{abstract}
The GlueX experiment is located at the Thomas Jefferson National Accelerator Facility (JLab) in Newport News, VA, USA. It features a hermetic 4$\pi$ detector with excellent tracking and calorimetry capabilities. Its \SI{9}{\GeV} linearly polarized photon beam is produced from the \SI{12}{\GeV} electron beam, delivered by JLab's Continuous Electron Beam Accelerator Facility (CEBAF), via bremsstrahlung on a thin diamond and is incident on a LH2 target. GlueX recently finished its first data taking period and published first results.\\
The main goal of GlueX is to measure gluonic excitations of mesons. These so-called hybrid or exotic mesons are predicted by Quantum Chromodynamics (QCD) but haven't been experimentally confirmed yet. They can have quantum numbers not accessible by ordinary quark-antiquark pairs which helps in identifying them using partial wave analysis techniques. The search for exotic mesons requires a very good understanding of photoproduction processes in a wide range of final states, one of them being $pK^{+}K^{-}$ which contains many meson and baryon reactions. The  $\Lambda$(1520) is a prominent hyperon resonance in this final state and is the subject of this presentation.\\
This talk will give an introduction to the GlueX experiment and show preliminary results for the photoproduction of the $\Lambda$(1520) hyperon. The measurement of important observables like the photon beam asymmetry and spin-density matrix elements will be discussed and an outlook to possible measurements of further hyperon states in the $pK^{+}K^{-}$ final state will be given.
\end{abstract}

\section{Introduction}
The main goal of GlueX is to establish the existence and measure the properties of exotic (hybrid) mesons. In a simple quark model picture these hadrons, which are not forbidden by QCD, can be thought of as a $q\bar{q}$ pair with an additional constituent gluon contributing to the quantum numbers of the meson. Because of the gluon these mesons can have quantum numbers which are forbidden for pure $q\bar{q}$ states. Therefore they are referred to as exotic and measuring these quantum numbers in a partial wave analysis is a smoking gun for a non-$q\bar{q}$ state.\\
Before GlueX can attempt to identify exotic partial waves it is important to gain an excellent understanding of the non-exotic contributions to the measured final states which will make up the vast majority. For this talk the focus lies on the measurement of the $\Lambda$(1520) in the $pK^{-}K^{+}$ final state. This final state contains many meson and baryon resonances and the $\Lambda$(1520) is one of the most prominent. It is an important background in many final states that contain $pK^{-}$, as e.g. the LHCb pentaquark candidates \cite{Aaij2015,LHCbcollaboration2019}, and a good understanding of its properties is very desirable.

\section{The GlueX detector}
\begin{figure}
	\centering
	\begin{minipage}{0.47\linewidth}
		\centering
		\includegraphics[width=\linewidth]{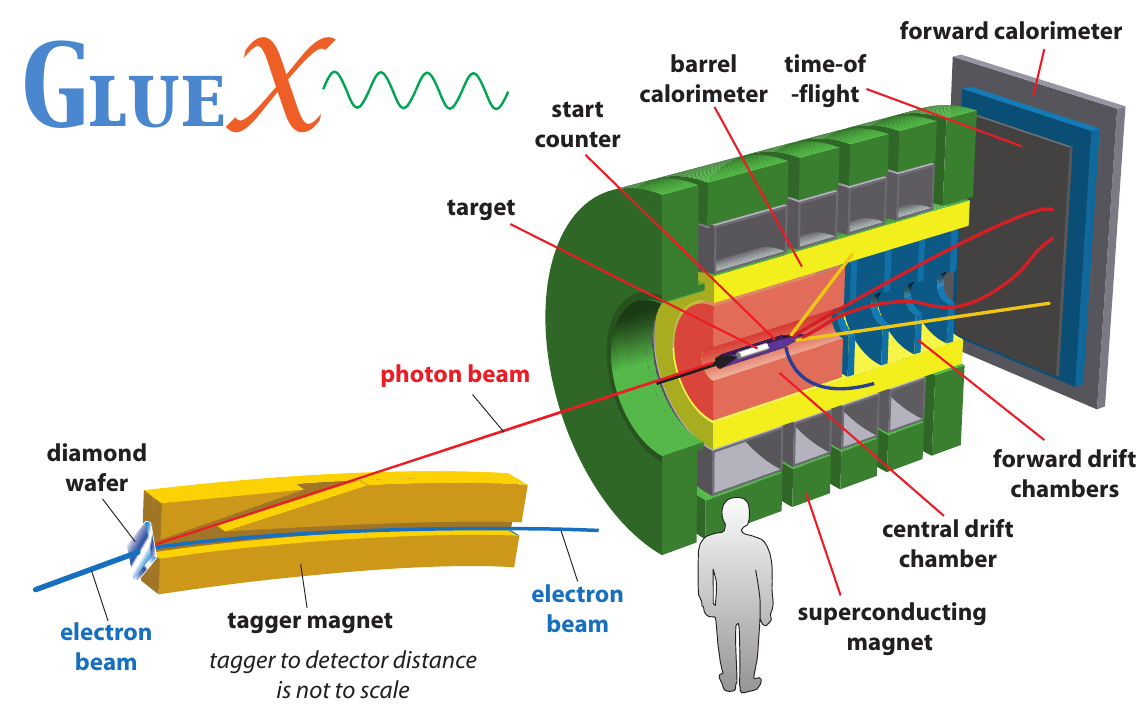}
		\caption{Overview over the GlueX detector setup. Taken from \cite{Collaboration2016}.}
		\label{fig:gluexoverview}
	\end{minipage}
	\hspace{0.5cm}
	\begin{minipage}{0.47\linewidth}
		\centering
		\includegraphics[height=6cm]{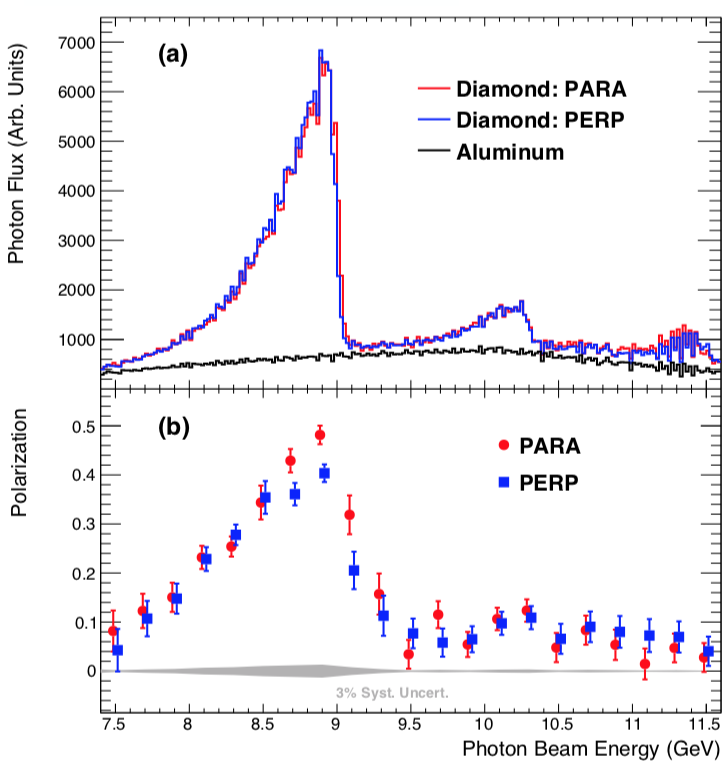}
		\caption{Photon flux (top) and polarization spectrum (bottom). The data was taken during the engineering run 2016 and is exemplary for the obtained spectra throughout the 2017 data taking period. Taken from \cite{AlGhoul2017}.}
		\label{fig:gluexbeam}
	\end{minipage}
\end{figure}

The GlueX experiment is situated in Hall D at the Continuous Electron Beam Accelerator Facility (CEBAF) at Jefferson Lab. An overview of the whole detector setup is shown in Figure \ref{fig:gluexoverview}. CEBAF provides an almost \SI{12}{\GeV} electron beam to the tagger hall of Hall D where it is converted to a photon beam via the bremsstrahlung process on a thin diamond. The resulting photon beam spectrum is shown in Figure \ref{fig:gluexbeam}. The flux (top) shows an enhancement at around \SI{9}{\GeV} which contains most of the polarized part (bottom) of the photon beam. The photon beam travels for about \SI{70}{\m} from the diamond to the detector where it gets collimated before it is incident on the LH2 target in the center of the detector setup. The target is surrounded by various drift chambers and calorimeters as well as a time-of-flight wall in the forward region. This combination of detectors results in an excellent coverage of the solid angle with very good energy and momentum resolution for both charged and neutral particles. The results shown in this talk are based on about $20\%$ of GlueX I data acquired in the spring 2017 beamtime. 

\section{$\Lambda$(1520) production}
\begin{figure}
	\centering
	\begin{tikzpicture}
		\node (img) {\includegraphics[width=0.8\linewidth]{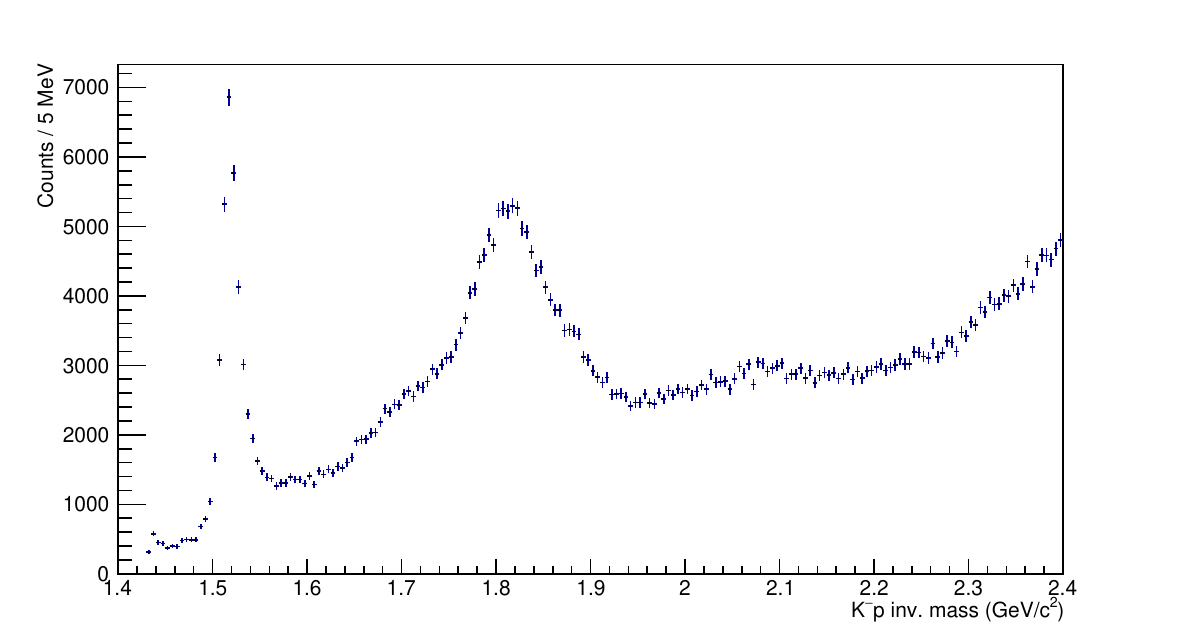}};
		\node [below left,text width=2cm,align=center,color=black] at (-1.9,1.9) {$\Lambda(1520)$};
		\node [below left,text width=5cm,align=center,color=black] at (2.8,-1.7) {various $\Lambda^{*}$ and $\Sigma^{*}$ baryons};
		\node[opacity=0.2,rotate=30] at (0,0)  {\includegraphics[width=0.35\linewidth]{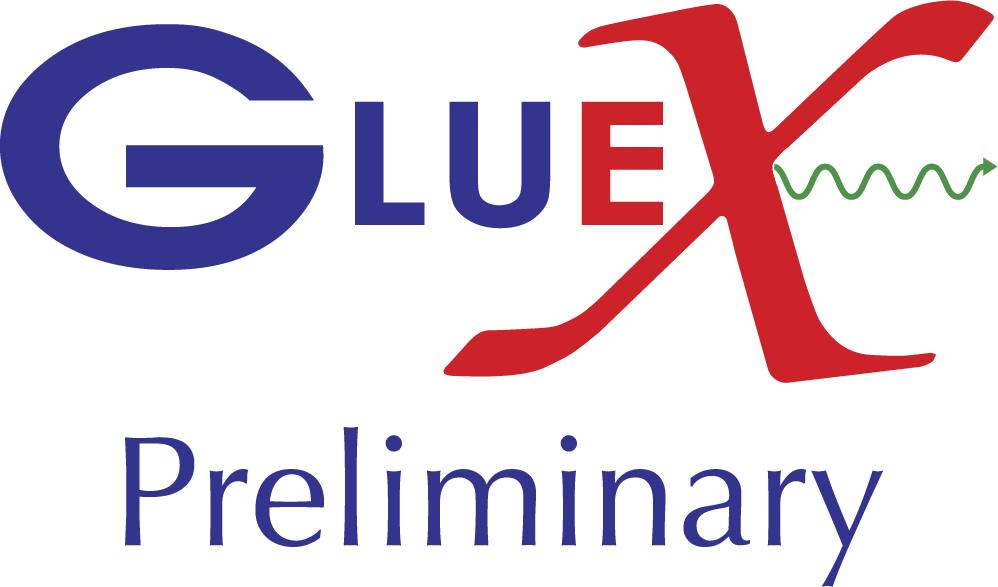}};
	\end{tikzpicture}
	\caption{$pK^{-}$ invariant mass spectrum after event selection. The $\Lambda$(1520) is visible as very sharp, prominent peak. At higher invariant masses one can see various heavier hyperons.}
	\label{fig:spectrum}
\end{figure}

The reaction of interest is $\gamma p \rightarrow K^{+}\Lambda(1520)\rightarrow K^{+}K^{-}p$. Due to GlueX's almost hermetic layout combined with very good energy and momentum resolution the reaction can be measured in an exclusive final state. The timing signals from the detectors are used to identify the proton and kaons and four-momentum conservation and vertex constraints are applied through a kinematic fit. The $pK^{-}$ invariant mass spectrum after event selection can be seen in Figure \ref{fig:spectrum}.
The $\Lambda$(1520) is visible as very prominent peak around \SI{1.52}{\GeV}. A cut is placed from \SI{1.46}{\GeV} to \SI{1.58}{\GeV} and the sPlot technique \cite{Pivk2005} is used to subtract the remaining background.

\subsection{Beam asymmetry}
\begin{figure}
	\centering
	\begin{tikzpicture}
		\node (img) {\includegraphics[trim={0 0.1cm 0 0},clip,width=0.7\linewidth]{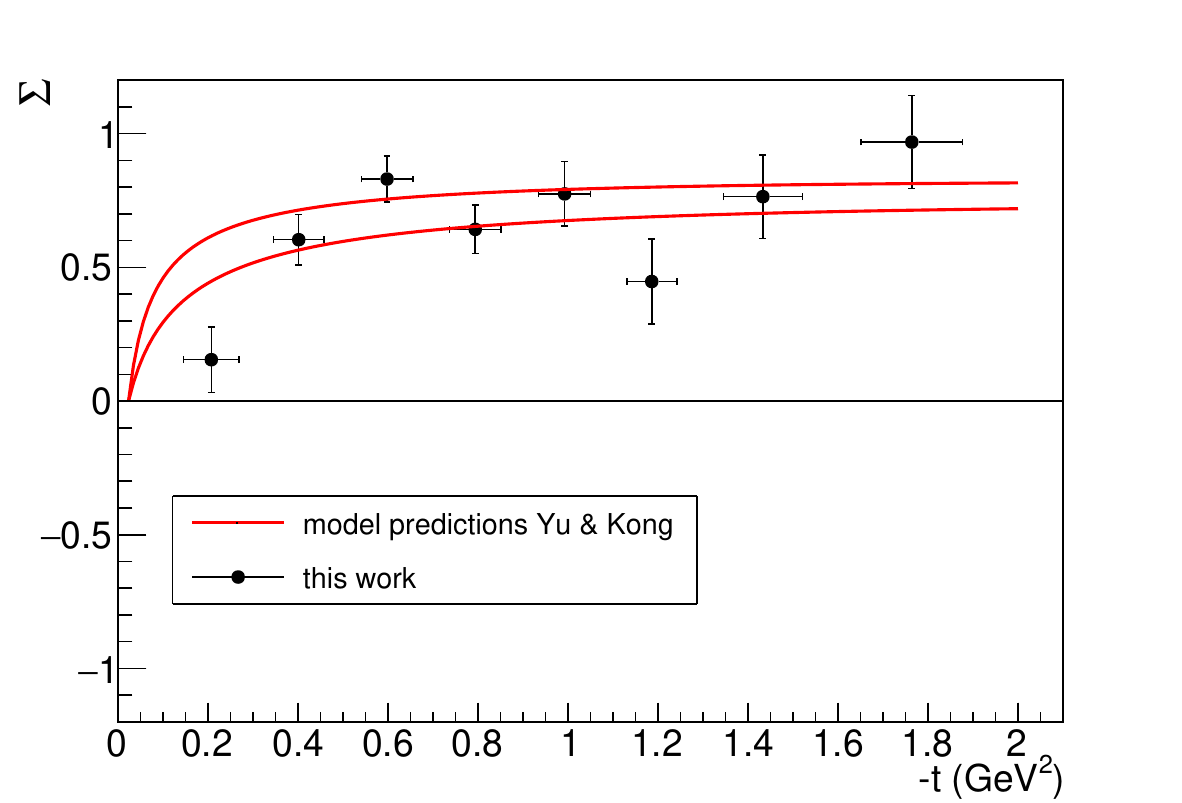}};
		\node[opacity=0.2,rotate=30] at (0,0)  {\includegraphics[width=0.4\linewidth]{GlueX_Preliminary.png}};
	\end{tikzpicture}
	\caption{Preliminary results for the beam asymmetry $\Sigma$ in the energy range of $E_{\gamma} =$ \SIrange[range-phrase = --]{8.2}{8.8}{\GeV}. The errors shown are statistical only. Red lines are model predictions (priv. communication based on \cite{Yu2017a}).}
	\label{fig:ba}
\end{figure}

The first observable to be extracted from the data is the beam asymmetry $\Sigma$ which contains some information on the production process of the reaction. It is defined as amplitude of the $\cos(2\phi)$ modulation of the production cross section, with a linearly polarised photon beam and an unpolarised target, in the laboratory frame. It can be extracted with an extended maximum likelihood fit with the likelihood function written as
\begin{align}
	\ln\mathcal{L} = \frac{\sum w_{i}}{\sum w^{2}_{i}}\left[\sum_{i}w_{i}\ln\mathcal{I}(\phi_{i,K^{+}}) - \int\text{d}\phi_{K^{+}}\,\mathcal{I}(\phi_{K^+}) \eta(\phi_{K^{+}})\right] \label{eq:loglikelihood}
\end{align}
and with the intensity function given by
\begin{align}
	\mathcal{I}(\phi_{K^{+}}) = 1 - P_{\gamma}\Sigma\cos\left(2\left(\phi_{K^{+}}-\phi_{0}\right)\right) \,.
\end{align}
$\phi_{K^{+}}-\phi_{0}$ is the angle between the outgoing $K^{+}$, which defines the production plane, and the polarisation plane of the beam photon and $P_{\gamma}$ is the degree of linear polarisation of the incoming photon beam. The $w_{i}$ denote the previously extracted sWeights taking account of the background under the $\Lambda$(1520). The fraction of sums-of-weights corrects the likelihood function such that the fit reports the correct errors (c.f. \cite{Aaij2015}). The acceptance $\eta$ is corrected for by approximating the integration with a sum over accepted simulated phase space Monte Carlo events. Preliminary results of the fit in bins of momentum transfer $-t = \left(p_{\gamma}-p_{K^{+}}\right)^{2}$ are shown in Figure \ref{fig:ba}.
The black points show the fit results with the vertical error bars showing the statistical error on the results and the horizontal error bars indicating the standard deviation of the $-t$-distribution within a bin. Note that systematic errors have not been determined yet for the measurement. The red lines show two calculations of the beam asymmetry by Yu and Kong based on their model described in \cite{Yu2017a}. The difference between them comes from two different data sets used to fix model parameters. In general the agreement between data and model is good.

\subsection{Spin-density matrix elements}
\begin{figure}
	\centering
	\begin{tikzpicture}
		\node (img) {\includegraphics[width=\linewidth]{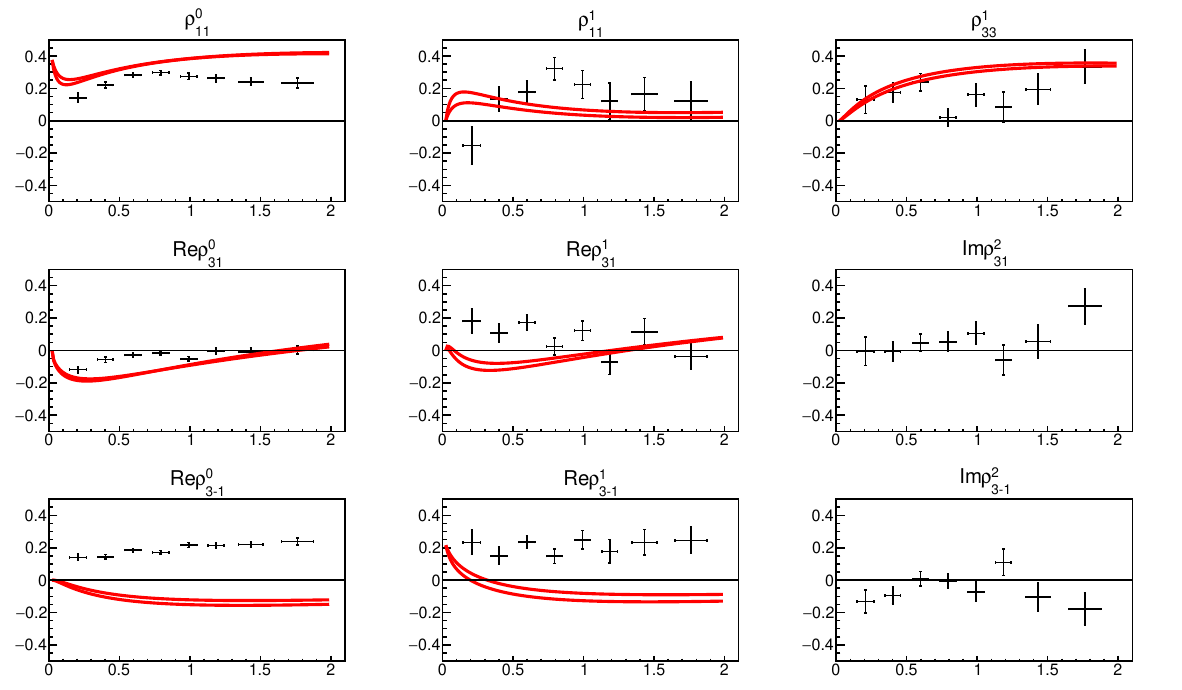}};
		\node[opacity=0.2,rotate=30] at (0,0)  {\includegraphics[width=.5\linewidth]{GlueX_Preliminary.png}};
		\draw[arrows=->,line width=.6pt](-6,-5)--(6,-5);
		\node at (4,-5.3) {\small$-t (\si{\GeV^{2}})$};
	\end{tikzpicture}
	\caption{Preliminary results for the nine independent SDMEs for $E_{\gamma} =$ \SIrange[range-phrase = --]{8.2}{8.8}{\GeV}. The errors shown are statistical only. Red lines are model predictions (priv. communication based on \cite{Yu2017a}).}
	\label{fig:sdme}
\end{figure}

\begin{figure}
	\centering
	\begin{tikzpicture}
      		\node (img) { \includegraphics[width=\linewidth]{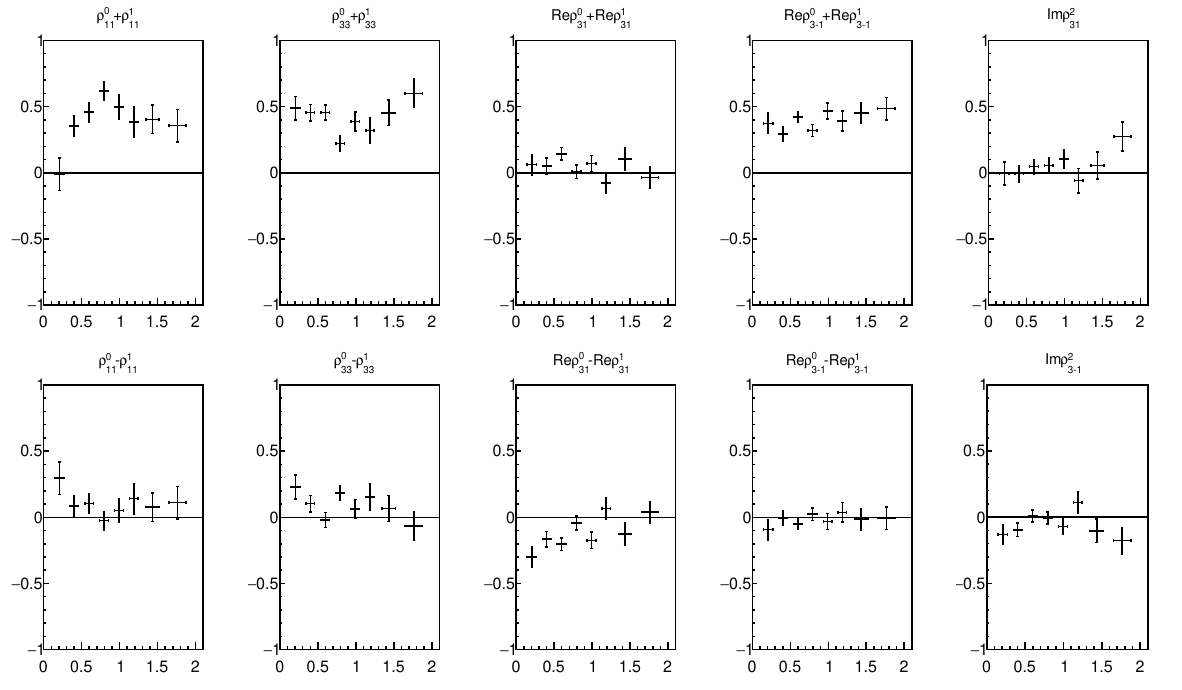}};
		\node [below left,text width=3cm,align=center,color=black,font=\boldmath] at (-4.8,1.3){\scalebox{.75}{\tiny{$|N_{0}|^{2}+|N_{1}|^{2}$}}};
		\node [below left,text width=3cm,align=center,color=black,font=\boldmath] at (-4.8,-3.4){\scalebox{.75}{\tiny{$|U_{0}|^{2}+|U_{1}|^{2}$}}};
		\node [below left,text width=3cm,align=center,color=black,font=\boldmath] at (-1.5,1.3){\scalebox{.75}{\tiny{$|N_{-1}|^{2}+|N_{2}|^{2}$}}};
   		\node [below left,text width=3cm,align=center,color=black,font=\boldmath] at (-1.5,-3.4) {\scalebox{.75}{\tiny{$|U_{-1}|^{2}+|U_{2}|^{2}$}}};
		\node [below left,text width=3cm,align=center,color=black,font=\boldmath] at (1.7,1.3){\scalebox{.75}{\tiny{Re($N_{-1}N_{0}^{*}-N_{1}N_{2}^{*}$)}}};
		\node [below left,text width=3cm,align=center,color=black,font=\boldmath] at (1.7,-3.4){\scalebox{.75}{\tiny{Re($U_{-1}U_{0}^{*}-U_{1}U_{2}^{*}$)}}};
		\node [below left,text width=3cm,align=center,color=black,font=\boldmath] at (4.9,1.3){\scalebox{.75}{\tiny{Re($N_{0}N_{2}^{*}+N_{1}N_{-1}^{*}$)}}};
		\node [below left,text width=3cm,align=center,color=black,font=\boldmath] at (4.9,-3.4) {\scalebox{.75}{\tiny{Re($U_{0}U_{2}^{*}+U_{1}U_{-1}^{*}$)}}};
		\node [below left,text width=3cm,align=center,color=black,font=\boldmath] at (8,1.3){\tiny{mix}};
		\node [below left,text width=3cm,align=center,color=black,font=\boldmath] at (8,-3.4) {\tiny{mix}};
		\node[opacity=0.2,rotate=30] at (0,0)  {\includegraphics[width=0.5\linewidth]{GlueX_Preliminary.png}};
		\draw[arrows=->,line width=.6pt](-6,-5.3)--(6,-5.3);
		\node at (5,-5.6) {\small$-t (\si{\GeV^{2}})$};
	\end{tikzpicture}
	\caption{Preliminary results for the combinations of SDMEs that can be related to purely natural or purely unnatural reaction amplitudes in the energy range of $E_{\gamma} =$ \SIrange[range-phrase = --]{8.2}{8.8}{\GeV}. The errors shown are statistical only.}
	\label{fig:sdmeinterpretation}
\end{figure}

To make most of the information available through the decay of the $\Lambda$(1520) spin-density matrix elements (SDMEs) are extracted. They provide means to parameterise the the angular decay distribution of the $\Lambda$(1520) in terms of nine independent real-valued numbers. A common frame of choice for the extraction of SDMEs is the Gottfried-Jackson frame \cite{Gottfried1964}, sometimes called t-channel helicity frame. In this frame the $\Lambda$(1520) is at rest, the z-axis is chosen anti-parallel to the incoming target proton, the y-axis is the cross-product of incoming photon beam direction and outgoing $K^{+}$ and the x-axis is chosen as cross-product of y-axis and z-axis. The angular distribution can then be written as
\begin{align}
	W_0(\varphi,\theta) &= \frac{1}{4\pi}\left[3\left(\frac{1}{2}-\rho^0_{11}\right)\sin^2(\theta) + \rho^0_{11}\left(1+3\cos^2(\theta)\right)\right. \nonumber\\
		& \hspace{1cm}\left. - 2\sqrt{3}\left(\text{Re}(\rho^0_{31})\cos(\varphi)\sin(2\theta) + \text{Re}(\rho^0_{3-1})\cos(2\varphi)\sin^2(\theta)\right)\vphantom{\frac{1}{2}}\right]\nonumber\\
	W_1(\varphi,\theta) &= \frac{1}{4\pi}\left[\vphantom{\left(1+3\cos^2(\theta)\right)^{2}}3\rho^1_{33}\sin^2(\theta) + \rho^1_{11}(1+3\cos^2(\theta))\right. \nonumber\\
		& \hspace{1cm}\left. - 2\sqrt{3}\left(\text{Re}(\rho^1_{31})\cos(\varphi)\sin(2\theta) + \text{Re}(\rho^1_{3-1})\cos(2\varphi)\sin^2(\theta)\right)\right] \nonumber\\
	W_2(\varphi,\theta) &= \frac{1}{4\pi}\left[2\sqrt{3}\left(\text{Im}(\rho^2_{31})\sin(\varphi)\sin(2\theta) + \text{Im}(\rho^2_{3-1})\sin(2\varphi)\sin^2(\theta)\right)\right] \nonumber\\
	~\nonumber\\ 
	W(\varphi,\theta,\Phi)   &= W_0(\varphi,\theta) - P_\gamma\cos(2\Phi)W_1(\varphi,\theta) - P_\gamma\sin(2\Phi)W_2(\varphi,\theta) \label{eq:sdme}
\end{align}
with $\rho_{2m2m'}$ being the SDMEs, $\varphi$ and $\theta$ being the angles of the $K^{-}$ in the Gottfried-Jackson frame and $\Phi$ being the angle between production plane and polarisation plane in the laboratory frame. In order to extract the SDMEs from the data a likelihood function as in Eq \eqref{eq:loglikelihood} is defined with the intensity function $\mathcal{I}(\phi_{K^{+}})$ being replaced by $W(\varphi,\theta,\Phi)$ from Eq \eqref{eq:sdme}. Due to the high complexity of this fit function (nine parameters and three variables) a Markov Chain Monte Carlo (MCMC) method is employed where the parameter space is explored numerically through a Metropolis-Hastings algorithm. The final results are then reported as mean and standard deviation of the resulting parameter distributions. The preliminary results are shown in Figure \ref{fig:sdme}.
As before the black crosses show the fit results with horizontal error bars indicating the standard deviation of the momentum transfer distribution within a bin. The vertical error bars are the standard deviations of the parameter distributions within the Markov Chain. The red lines come from the same model calculation as shown for the beam asymmetry. It is interesting to see that some of the fit results show good agreement but others seem to have a different sign in the calculation or even a slightly different shape. This shows that these measurements will have an impact on the current theoretical descriptions available.\\
In order to interpret the SDME results one can write the reaction amplitudes in the reflectivity base. For high energies and t-channel production the natural ($N$) and unnatural ($U$) exchange amplitudes can be separated when the beam is linearly polarised. A natural exchange amplitude means a vector or tensor meson exchange and an unnatural exchange amplitude means a pseudoscalar or axial-vector exchange. These amplitudes can be directly related to combinations of SDMEs. The combinations of SDMEs and matching amplitudes of purely natural or unnatural exchange can be seen in Figure \ref{fig:sdmeinterpretation}.
These preliminary results show that unnatural amplitude combinations are mostly compatible with zero while natural amplitude combinations have mostly positive values. This indicates that natural exchanges, such as vector mesons, dominate over unnatural exchanges, such as pseudo-scalars, in this reaction.

\section{Summary and outlook}
We have shown preliminary results for the beam asymmetry $\Sigma$ and spin-density matrix elements for the reaction $\gamma p\rightarrow K^{+}\Lambda(1520)\rightarrow K^{+}K^{-}p$ at photon energies of $E_{\gamma} =$ \SIrange[range-phrase = --]{8.2}{8.8}{\GeV}. These measurements are the first for this reaction in this energy range and will help to constrain theoretical models describing the production of this excited hyperon. Effort is currently under way to also provide measurements for (differential) cross-sections. Combined with the current results, these will provide further information to constrain the theoretical model of processes contributing to the $\Lambda$(1520) photoproduction reaction and hence improve our understanding of the production of standard and exotic states.

\ack
This work was supported by the Scottish Universities Physics Alliance. This material is based upon work supported by the U.S. Department of Energy, Office of Science, Office of Nuclear Physics under contracts DE- AC05-06OR23177. We thank V. Mathieu for providing us with the expressions relating SDMEs to amplitudes in reflectivity base.

\section*{References}
\bibliographystyle{iopart-num}
\bibliography{proceedings.bbl}

\providecommand{\newblock}{}
\begin{thebibliography}{1}
\expandafter\ifx\csname url\endcsname\relax
  \def\url#1{{\tt #1}}\fi
\expandafter\ifx\csname urlprefix\endcsname\relax\def\urlprefix{URL }\fi
\providecommand{\eprint}[2][]{\url{#2}}

\bibitem{Aaij2015}
  Aaij R {\em et al.}
  2015
  {\em Phys. Rev. Lett. \/} {\bf 115} 072001

\bibitem{LHCbcollaboration2019}
  Aaij R {\em et al.}
  2019
  {\em Phys. Rev. Lett. \/} {\bf 122} 222001

\bibitem{Collaboration2016}
  {Al Ghoul} H {\em et al.}
  2016
  {\em AIP Conference Proceedings\/}
  (AIP Publishing LLC)
  vol 1735 020001

\bibitem{AlGhoul2017}
  {Al Ghoul} H {\em et al.}
  2017
  {\em Phys. Rev. C\/} {\bf 95} 042201

\bibitem{Pivk2005}
  Pivk M and {Le Diberder} F~R
  2005
  {\em Nucl. Instrum. Meth. A\/} {\bf 555} 356--369

\bibitem{Yu2017a}
  Yu B~G and Kong K~J
  2017
  {\em Phys. Rev. C\/} {\bf 96} 025208

\bibitem{Gottfried1964}
  Gottfried K and Jackson J~D
  1964
  {\em Il Nuovo Cimento\/} {\bf 33} 309--330

\end{thebibliography}

\end{document}